\documentclass[aps,twocolumn,floats,nofootinbib,prx]{revtex4}
\usepackage{graphics,graphicx,epsfig}
\usepackage{amssymb,color}
\usepackage{amsmath}
\usepackage{epsf,epstopdf,wrapfig}
\usepackage[hidelinks]{hyperref}

\newcommand{\beq}{\begin{equation}}
\newcommand{\eeq}{\end{equation}}
\newcommand{\beqn}{\begin{eqnarray}}
\newcommand{\eeqn}{\end{eqnarray}}

\begin{document}

\title{Finding the last bits of positional information}

\author{Lauren McGough,$^{a,b}$ Helena Casademunt,$^{c}$ Milo\v{s} Nikoli\' c,$^a$ Mariela D.~Petkova,$^d$ Thomas Gregor,$^{a,e}$ and William Bialek,$^{a}$}

\affiliation{$^a$Joseph Henry Laboratories of Physics and Lewis--Sigler Institute for Integrative Genomics, Princeton University, Princeton NJ 08544 USA\\
$^b$Department of Ecology and Evolution, The University of Chicago, Chicago IL 60637\\
$^c$Department of Physics and $^d$Program in Biophysics, Harvard University, Cambridge MA 02138\\
$^e$Department of Developmental and Stem Cell Biology UMR3738, Institut Pasteur, 75015 Paris, France}

\date{\today}
\begin{abstract}
In a developing embryo, information about the position of cells is encoded in the concentrations of ``morphogen'' molecules.  In the fruit fly, the local concentrations of just a handful of proteins encoded by the gap genes are sufficient to specify position with a precision comparable to the spacing between cells along the anterior--posterior axis.  This matches the precision of downstream events such as the striped patterns of expression in the pair-rule genes, but is not quite sufficient to define unique identities for individual cells. We demonstrate theoretically that  this  information gap  can be bridged if positional errors are spatially correlated, with relatively long correlation lengths. We then show experimentally that these correlations are present, with the required strength, in the fluctuating positions of the pair-rule stripes, and this can be traced back to the gap genes.  Taking account of these correlations, the available information matches the information needed for unique cellular specification, within error bars of $\sim 2\%$.  These observation support a precisionist view of information flow through the underlying genetic networks, in which accurate signals are available from the start and preserved as they are transformed into the final spatial patterns.
\end{abstract}

\maketitle

\section{Introduction}

\begin{figure}[b!]
\centerline{\includegraphics[width = 0.9\linewidth]{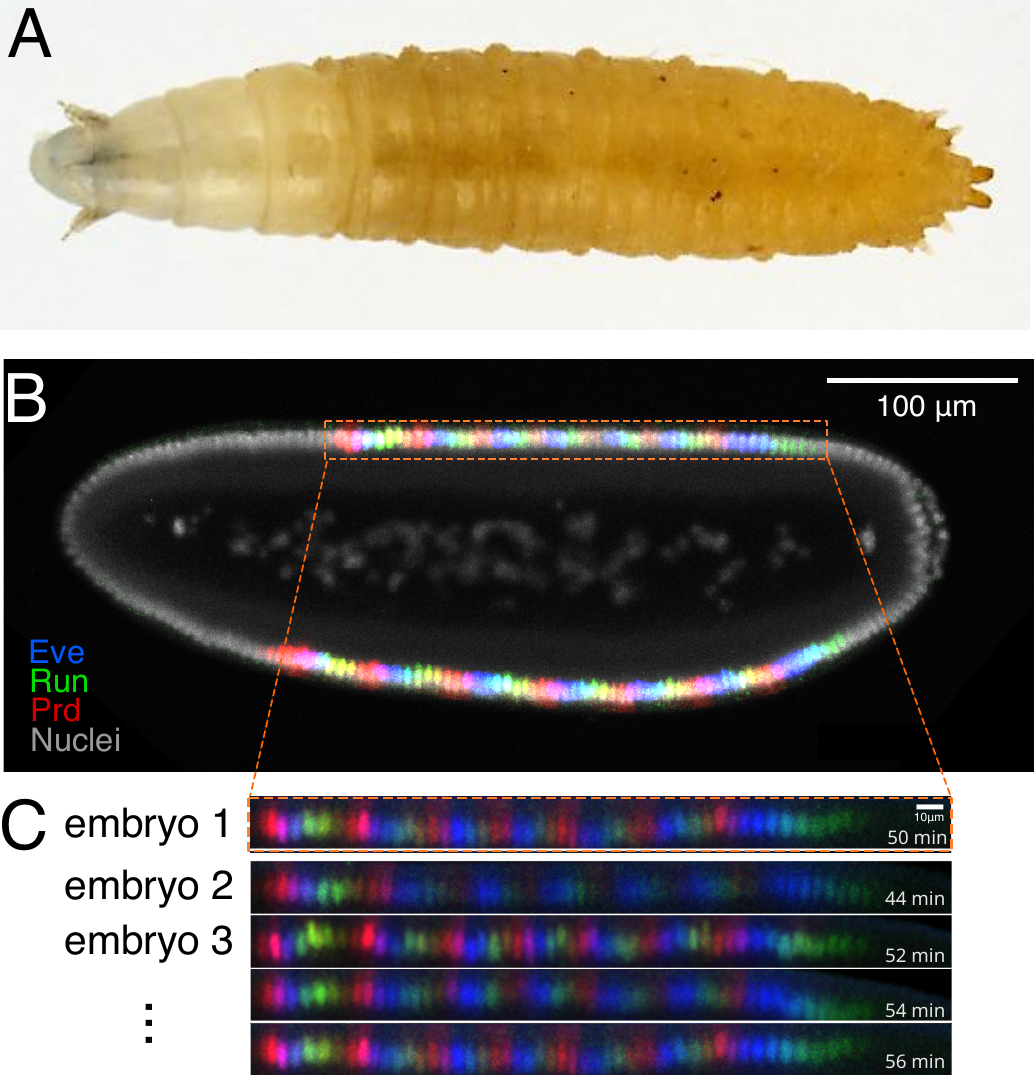}}
\caption{Segmented {\em Drosophila} body plan.  (A) Brightfield color image of a $5\,$mm long $3^{\rm rd}$ instar larva of the fruit fly {\em Drosophila melanogaster} \cite{bugs} with clearly visible segments.  (B) An optical section through an embryo stained for three of the ``pair-rule'' proteins, $50\,$min into nuclear cycle 14 ($\sim 3\,{\rm h}$ after oviposition), showing striped patterns that align with the body segments; data from Ref \cite{petkova+al_19}.  (C) As in (B), from multiple embryos, illustrating the pattern reproducibility. Time in nuclear cycle 14 indicated at bottom right of each profile.
\label{stripes}}
\end{figure}

During the development of an embryo, cell fates are determined in part by the concentrations of specific morphogen molecules that carry information about position  \cite{turing_52,wolpert_69,tkacik+gregor_21}.   For the early stages of fruit fly development, all of these molecules have been identified \cite{nusslein+wieschaus_80,lawrence_92,ew+cnv_16}.  For patterning along the main body axis, spanning from anterior to posterior (AP), information flows from primary maternal morphogens to an interacting network of gap genes to the pair-rule genes \cite{riverapomar_96,jaeger_11}, whose striped patterns of expression provide a precursor of the segmented body plan in the fully developed organism, visible within three hours after the egg is laid (Fig.~\ref{stripes}).  It has been known for some time that, at this stage in development, essentially every cell ``knows'' it's fate \cite{gergen+al_86}, so it is natural to ask how this information is encoded, quantitatively, in the concentrations of the relevant morphogens. 

The expression levels of the gap genes provide enough information to specify the positions of individual cells with an accuracy  $\sim 1\%$ of the embryo's length \cite{dubuis+al_13}. This matches the precision with which the stripes of pair-rule expression are positioned, and the precision of macroscopic developmental events such as the formation of the cephalic furrow  \cite{liu+al_13}.  Further, the algorithm that extracts optimal estimates of position from the expression levels of the gap genes also predicts, quantitatively, the distortions of the striped pattern in mutant flies with deletions of the maternal inputs \cite{petkova+al_19}.  At the moment when pair-rule stripes are fully formed, just before gastrulation, there are fewer than one hundred rows of cells along the length of the embryo, so it is tempting to think that positional signals with $1\%$ accuracy  define unique cellular identities.  In fact, this is not quite correct  \cite{dubuis+al_13}: if each cell makes independent positional errors drawn from a Gaussian distribution, then there is a small but significant probability that neighboring cells will get ``crossed signals,''  driving errors in cell fate determination.

The small difference between $1\%$ positional errors and unique cellular identities provides an interesting test case in the search for a more quantitative understanding of living systems.  In physics, we are used to the idea that small quantitative discrepancies can be signs of qualitatively new ideas or mechanisms.
But in complex biological systems one might worry that small discrepancies reflect experimental errors or over--simplifications in interpretation.  If correct, these concerns would limit our ambitions for quantitative theory in the physics tradition. However the small discrepancies need to be re--examined in light of dramatic improvements in experimental precision \cite{yu+al_06,taniguchi+al_10,dubuis+al_13b}.

Here we take the small quantitative discrepancy in positional information seriously.  On the theoretical side, we clarify the problem, defining an ``information gap,'' and show that this gap can be closed if errors in the positional signals are spatially correlated over relatively long distances.  Early work by Lott and colleagues \cite{lott+al_07} detected such correlations in mRNA levels of gap and pair-rule genes;  subsequent work found that noise in different combinations of protein levels in the gap gene network are correlated significantly over the entire length of the embryo \cite{krotov+al_14}.  On the experimental side we re--examine these correlations, measuring the positions of stripes in the concentrations of pair-rule proteins.  We find that the extent of these correlations is what is needed to close the information gap between positional errors and unique cellular identities, quantitatively.

\section{Defining the problem}
\label{sec:problem}

In the early fly embryo, cells have access to the concentrations of morphogens, and these concentrations are continuously graded.  From these concentrations, it is possible to decode an estimate of position, which we label as $\hat x_{\rm n}$ in cell $\rm n$ \cite{petkova+al_19}.  We expect that these estimates are correct on average, so that $\langle \hat x_{\rm n} \rangle = {\rm n}L /N$, where there are $N$ cells along the length $L$ of the embryo.\footnote{For simplicity we imagine that the problem is one--dimensional so that cells need to know their position only along one axis.  In the early fly embryo, patterning signals along the two major axes are largely independent \cite{nusslein-volhard_89, nusslein-volhard_91}, justifying this simplification.}  
However the signals are noisy, so decoding in one cell will have errors,
\begin{eqnarray}
\hat x_{\rm n} &=& {\rm n}L /N + \delta x_{\rm n}, \label{est1}\\
\langle (  \delta x_{\rm n} )^2 \rangle &=& \sigma_x^2 .\label{est2}
\end{eqnarray}
For simplicity, but guided by the experimental observations \cite{dubuis+al_13,tkacik+al_15,petkova+al_19}, we assume that $\sigma_x$ is the same for all cells and that the distribution of $\delta x_{\rm n}$ is Gaussian.  Here we are interested in the question of whether cells get signals that define the correct ordering along the axis so that $\hat x_{\rm n+1} > \hat x_{\rm n}$ for all cells, or whether they can get ``crossed signals'' such that $\hat x_{\rm n+1} < \hat x_{\rm n}$.  

\begin{figure}
\centerline{\includegraphics[width = \linewidth]{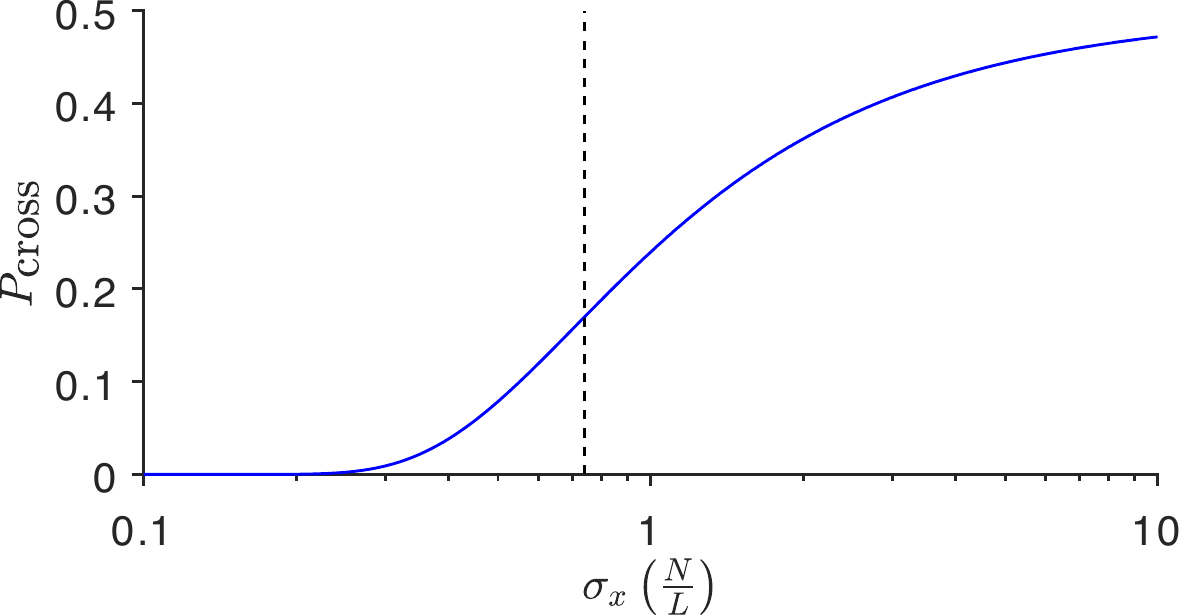}}
\caption{Probability of ``crossed signals'' between two neighboring cells as a function of the positional error, assuming that noise is independent in each cell [Eq (\ref{eq-pcross})]. Dashed vertical line marks the experimental value of positional noise, $\sigma_x\sim 0.01 L$, which corresponds to less than the mean distance between neighboring cells $L/N$ \cite{dubuis+al_13}.
\label{Pcross}}
\end{figure}

If we look at two neighboring cells,  then the probability of incorrect ordering is
\begin{equation}
P_{\rm cross} \equiv {\rm Pr}(\hat x_{\rm n+1} < \hat x_{\rm n}).
\end{equation}
To find the probability of a wrong ordering we can take a look at the distribution of the distance to the next cell $y= \hat x_{\rm n+1} - \hat x_{\rm n}$. But since $\hat x_{\rm n+1}$ and $\hat x_{\rm n}$ both are Gaussian, their difference $y$ is also Gaussian, with mean equal to $\langle y\rangle = L/N$.  If the noise is independent in each cell, then the variance of this difference signal will be 
$\langle (\delta y)^2 \rangle = 2\sigma_x^2$.
Incorrect ordering happens when $y<0$, which then has probability
\begin{eqnarray}
P_{\rm cross} &=& \int_{-\infty}^0 {{dy}\over{\sqrt{4\pi\sigma_x^2}}}
 e^{-(y-L/N)^2/4\sigma_x^2}\\
 &=& {1\over\sqrt{4\pi}} \int_{1/z} ^\infty dx \,e^{-x^2/4},
\label{eq-pcross}
\end{eqnarray}
with $z = \sigma_x (N/L)$, as shown in Fig.~\ref{Pcross}.
If positional errors are comparable to the spacing between cells, $\sigma_x \sim L/N$, the probability of an error is nearly 24\%; for the experimental value $\sigma_x \sim (0.74) L/N$ \cite{dubuis+al_13}, crossed signals will occur in $\sim 16\%$ of cells. With $N\sim 74 \pm5$ rows of cells along the AP axis \cite{dubuis+al_13}, the probability that all signals come in the right order would be vanishingly small.\footnote{This uncertainty in $N$ may seem large, but what will matter below is  the information required to specify unique cellular identities, $I_{\rm unique} = \log_2 N$.  Although $\delta N/N$ is nearly ten percent, $\delta I_{\rm unique}/I_{\rm unique}$ is less than two percent.}

This failure to specify unique cellular identities can be given a simple information-theoretic interpretation.  To specify one cell uniquely out of $N$ requires $I_{\rm unique} = \log_2 N\,{\rm bits}$ of information \cite{shannon_48,bialek_12}.  On the other hand, if we have signals that represent a continuous position $x$ drawn uniformly from the range $0 < x \leq L$, 
and these signals have Gaussian noise with (small) standard deviation $\sigma_x$, as described above, then the amount of information the signal conveys about position is
\begin{equation}
I_{\rm position} = \log_2 L - \log_2\left( \sqrt{2\pi e} \sigma_x\right),
\label{Ipos1}
\end{equation}
where the first term is the entropy of the uniform distribution of positions and the second term is the entropy of the Gaussian noise distribution \cite{bialek_12}.  Combining these we can define an ``information gap''
\begin{equation}
I_{\rm gap} \equiv  I_{\rm unique}- I_{\rm position} = \log_2 \left( {{N\sigma_x}\over L} \sqrt{2\pi e}\right) .
\end{equation}
As discussed below, we obtain a more accurate estimate of the information gap by averaging over measurements of $\sigma_x$ at multiple points along the embryo, defined by the pair rule stripes, and we find $I_{\rm gap} = 1.39\pm 0.08 \,{\rm bits}$ (Appendix \ref{one-body}). Importantly this gap is measured per cell: it is not that the embryo is missing $\sim1.4\,{\rm bits}$ of information, but rather that {\em every cell} is missing this information.

\section{Extra information from correlations: Theory}
\label{sec-theory}

In order to address this information gap directly, we leverage the concept that correlated noise facilitates enhanced information transmission. While correlated noise is typically viewed as challenging due to its resistance to averaging, in the context of neighboring cells making correlated errors in position, it mitigates the probability of receiving ``crossed signals,'' as previously defined. Here we develop these considerations more formally.

Information is roughly the difference in entropy between the signal and the noise, where entropy measures the (log) volume in phase space that is occupied by a set of points.  When random variables become correlated, the volume and hence the entropy is reduced, even if the variances of the individual variables are unchanged.  In our example, with correlations, the full pattern of points $\{\hat x_1,\, \hat x_2,\, \cdots,\, \hat x_N\}$ fills a smaller volume in the space  $ [0, L]^N $ of possible positions for all the cells, and thus the embryo as a whole has access to more positional information.

More formally, we can define the correlation matrix $C$,
\begin{equation}
\langle   \delta x_{\rm n}  \delta x_{\rm m}  \rangle = \sigma_x^2 {C}_{\rm nm} ,
\end{equation}
with diagonal elements $C_{\rm nn} = 1$.  Assuming again that the noise $\delta x_{\rm n}$ is Gaussian, the reduction in noise entropy for the entire set of variables $\{\delta x_{\rm n}\}$ is given by the determinant of this matrix \cite{bialek_12},
\begin{equation}
\Delta S = - \frac{1}{2}\log_2\det C \,\,\,\,{\rm bits} ,
\label{eq-DeltaS}
\end{equation}
and this reduction in entropy is the gain in information.  Entropy is an extensive quantity, so that when $N$ is large the information gain per cell $I_{\rm extra} = \Delta S /N$ is finite.  Can $I_{\rm extra}$ be large enough to compensate for the information gap $I_{\rm gap}$?

\begin{figure}
\centerline{\includegraphics[width = \linewidth]{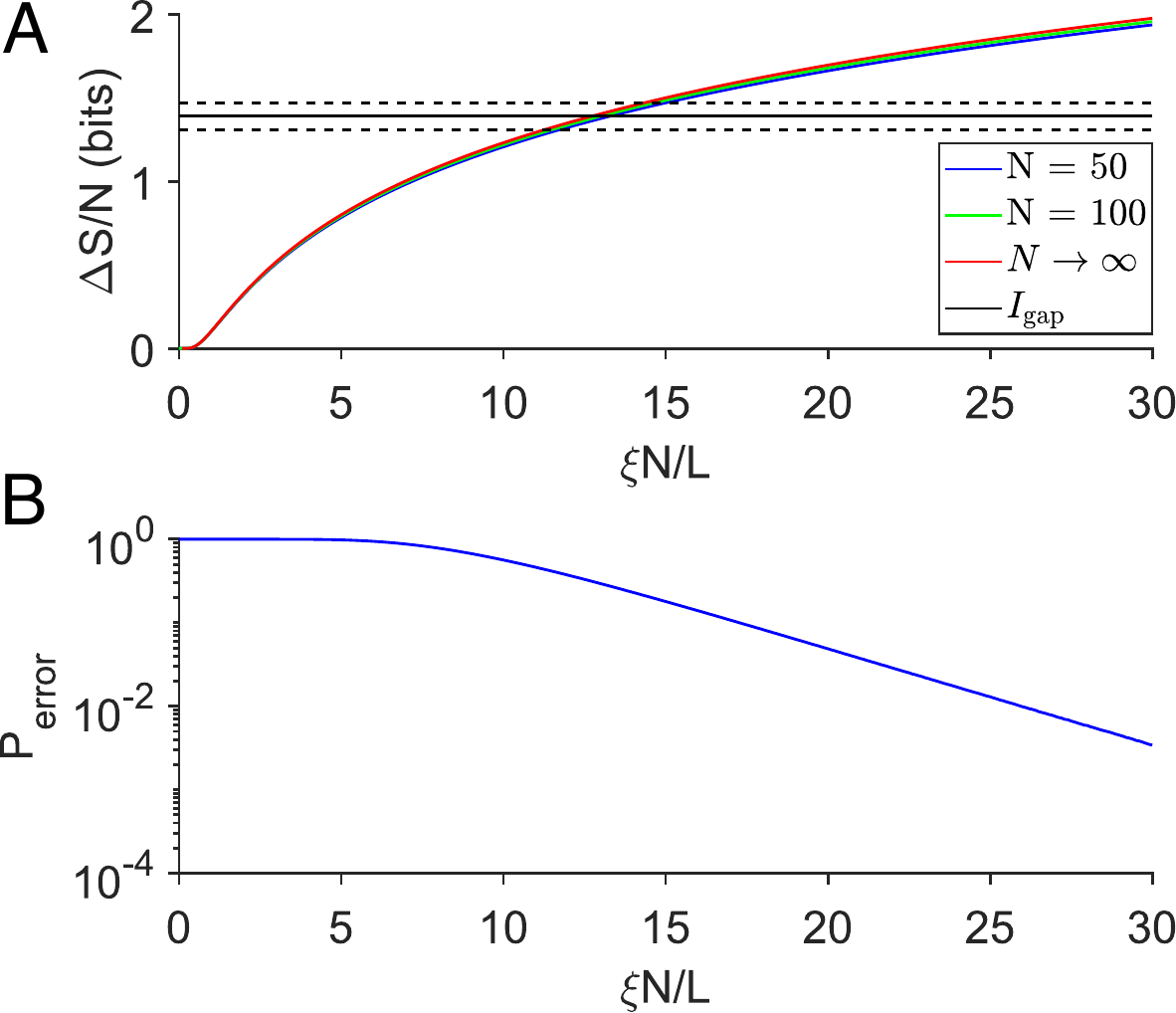}}
\caption{Extra information from correlations, as a function of the correlation length.  (A) Numerical results for  $N=50$ and $N=100$ from Eq (\ref{eq-DeltaS}) with the correlation matrix in Eq (\ref{C_exp});  analytic results for $N\rightarrow\infty$ from Eq (\ref{dS_inf}). Compare with the information gap from Appendix A (solid black line bracketed by dashed error bars).
(B) Probability of at least two signals being ``crossed,'' $\hat x_{{\rm n}+1} < \hat x_{\rm n}$ in a line of $N=74$ cells, with $\sigma_x/L = 0.01$.
\label{Pcross+DeltaS}}
\end{figure}

We expect that the correlation between fluctuations of positional signals in different cells depends on their spatial separation.  Then $C_{\rm nm}$ is a function of the distance between cells $\rm n$ and $\rm m$,  $d_{\rm nm} = |{\rm n} - {\rm m}|L/N$.   A natural functional form is an exponential decay of correlations,
\begin{equation}
C_{\rm nm} = e^{-d_{\rm nm}/\xi} ,
\label{C_exp}
\end{equation}
with correlation length $\xi$.  This is  what we would see if signals were encoded in the gradient of a single molecular species that has a lifetime $\tau$ and diffusion constant $D$, with $\xi = \sqrt{D\tau}$.  Although this is over--simplified, it is useful for building intuition about how the range of correlations determines the additional information.  Within this model it is straightforward to evaluate $\Delta S$ numerically, with results shown in Fig.~\ref{Pcross+DeltaS}A.

We can also give an analytic theory for $\Delta S$ in the large $N$ limit, leading to Eq (\ref{dS_inf}) and the red line in Fig.~\ref{Pcross+DeltaS}. If we define eigenvalues and eigenvectors of the matrix $C_{\rm nm}$,
\begin{equation}
\sum_{\rm m} C_{\rm nm} \phi_{\rm m}^\mu = \lambda_\mu \phi_{\rm n}^\mu ,
\end{equation}
then we have 
\begin{equation}
\Delta S = - \frac{1}{2}\sum_\mu \log_2 \lambda_\mu \,\,\,\,{\rm bits} .
\label{eq-DeltaS2}
\end{equation}
In the limit of large $N$ at fixed $N/L$, the ends of the embryo are  far away, and there is an effective translation invariance.  This means that the eigenvectors $\phi_{\rm n}^\mu$ are complex exponentials, $\phi_{\rm n}^\mu \propto \exp(iq_\mu {\rm n})$, or equivalently that the matrix $C_{\rm nm}$ is diagonalized by a discrete Fourier transform;\footnote{The discreteness is important.  If we take a continuum limit, so that the sum in Eq (\ref{lambda_q}) becomes an integral, the calculation is a bit simpler but leads to a significant over--estimate of $\Delta S$, even at large values of $\xi N/L$. } allowed values of $q_\mu$ are in the interval $-\pi \leq q < \pi $.  Then as $N\rightarrow\infty$ we find  the eigenvalues 
\begin{equation}
\lambda(q) \rightarrow \sum_{{\rm n}=-\infty}^\infty e^{-|{\rm n}|L/N\xi} e^{iq{\rm n}} = {{\sinh(L/N\xi)}\over{\cosh(L/N\xi) - \cos(q)}} ,
\label{lambda_q}
\end{equation}
and the change in entropy
\begin{eqnarray}
\Delta S/N &\rightarrow& 
- {1\over 2}\int_{-\pi}^\pi {{dq}\over{2\pi}} \log_2 \lambda (q) \\
&=& -{1\over 2}\log_2 \left[\frac{2 \sinh(L/N\xi)}{\sinh(L/N\xi)+\cosh(L/N\xi)}\right].
\label{dS_inf}
\end{eqnarray}
In Fig.~\ref{Pcross+DeltaS}A we see that this analytic result agrees with numerical results at $N=50$ and $N=100$, which agree with one another,  confirming that the fly embryo is large enough for the entropy to be extensive. We conclude that an information gap of $\sim$1.4 bits can be closed if correlations extend over distances $\xi \sim 13 (L/N)  \sim 0.18 L$.  
Lott and colleagues saw significant correlations across this range of distances for all the genes that they probed \cite{lott+al_07}, and combinations of gap gene protein levels have even longer correlation lengths \cite{krotov+al_14}.

Beyond the perhaps abstract information theoretic measures, we can evaluate the probability that all cells receive signals that are in the correct order, that is 
$\hat x_{\rm n+1} > \hat x_{\rm n}$ for all ${\rm n} = 1,\, 2\, \cdots ,\, N$. If correlations extend over a distance $\xi \sim 13 (L/N)$, then proper ordering will occur in more than $99\%$ of embryos, as illustrated in Fig.~\ref{Pcross+DeltaS}B.

\section{Extra information from correlations: Experiment}
\label{sec-exp}

Taking the information gap seriously, we {\em predict} that the noise in positional signals should be correlated over distances $\xi \sim 0.2 L$.  These distances are long compared to the separation between neighboring cells.  The first indication that such correlations exist came from experiments marking the boundaries of gene expression domains as seen through measurements of mRNA for selected gap genes and the pair rule gene {\em eve} \cite{lott+al_07}.   At the same time, it was reported that fluctuations in the concentration of a single gap gene product protein are correlated only over short distances \cite{gregor+al_07b}.   Analyzing simultaneous measurement on protein concentrations of four gap genes demonstrated that different combinations  or modes of the network have different correlation lengths \cite{krotov+al_14}; the longest correlation lengths are a significant fraction of the length of the embryo.   Finally, early analyses showed that errors in relative position are smaller than errors in absolute position \cite{dubuis+al_13}. All of this suggests that the noise in positional signals is spatially correlated. Can we make this statement more quantitative?

We analyze the experiments in Ref \cite{petkova+al_19}, which used immunofluorescence stainings to measure spatial profiles of protein concentration for three of the pair-rule genes {\em eve}, {\em prd}, and {\em rnt}  (Fig.~\ref{stripes}).  The data include $N_{\rm em} = 109$ embryos, fixed and stained in the time window from 35 to 60$\,$min after the start of nuclear cycle 14.   This is the period of cellularization, and as in previous work, the progress of the cellularization membrane provides a time marker with an accuracy of up to one minute \cite{dubuis+al_13b}.  For each of the three genes, the seven peaks in the striped concentration profile can be found automatically, and their locations vary linearly with time throughout this period \cite{antonetti+al_18}.  If we don't correct for this systematic dynamical behavior, the variance of stripe positions will be large and their fluctuations will be correlated, artificially.  We consider the noise in position to be the deviation from the best fit linear relation for each individual stripe marker.  The standard deviations then are consistently slightly below $\sigma_x \sim 0.01L$, and the distribution of fluctuations is well approximated by a Gaussian.  These results agree with previous work \cite{dubuis+al_13,petkova+al_19,antonetti+al_18}, and are summarized in Appendix \ref{one-body}.

Before analyzing correlations, we can use these data to make a more precise estimate of the information gap. If each cell has access to a positional signal with errors $\sigma_x ({\rm n})$, that might vary with $\rm n$, the average positional information available to a single cell is
\begin{equation}
I_{\rm position} = \log_2 L - {\bigg\langle} \log_2\left[ \sqrt{2\pi e} \sigma_x ({\rm n}) \right] {\bigg\rangle}_{\rm n} ,
\label{Ipos2}
\end{equation}
where $\langle \cdots \rangle_{\rm n}$ denotes an average over cells, generalizing Eq (\ref{Ipos1}).  Rather than making inferences about single cells, we have direct access to the signals that mark the locations of the stripes in the expression of three pair-rule genes, for a total of 21 features spread across half the AP axis.  The mean separation between the nearest stripes is $\Delta\bar{x} = 0.023 L$, just a few times larger than the spacing between cells.  Rather than introducing a model that would interpolate, we take the stripe positions themselves as the signals $x_{\rm n}$, now with ${\rm n} = 1,\, 2\, \cdots ,\,  21$, and the average in Eq (\ref{Ipos2}) becomes an average over stripes.

The challenge in evaluating the positional information is that random errors in our estimates of the errors $\sigma_x({\rm n})$ become systematic errors in estimates of information.  This problem of systematic errors was appreciated in the very first efforts to use information theoretic concepts to analyze biological experiments \cite{miller_55}.  The analysis of neural codes has been an important testing ground for methods to address these errors \cite{panzeri+treves_95,strong+al_98,paninski_03}; for a review see Appendix A.8 of Ref \cite{bialek_12}.   The approach we take here uses the fact that naive entropy estimates depend systematically on the size of the sample;  if we can detect this systematic dependence then we can extrapolate to infinite data, as described in Appendix \ref{one-body}.  The result is that $I_{\rm gap} = 1.39 \pm 0.08\,{\rm bits/cell}$.

The idea of positional information is that cells have access to a signal that represents position along the axis of the embryo \cite{wolpert_69,tkacik+al_15}.  In the discussion above we have taken this idea very seriously, identifying the signal in each cell as $\hat x_{\rm n}$.  But the signals we observe are the positions of stripes in three different pair-rule genes, and the different stripes for each gene are controlled by different enhancers responding to distinct combinations of transcription factors.  We need to test the hypothesis that these multidimensional molecular concentrations encode a single positional variable.

\begin{figure}[t]
\centerline{\includegraphics[width = \linewidth]{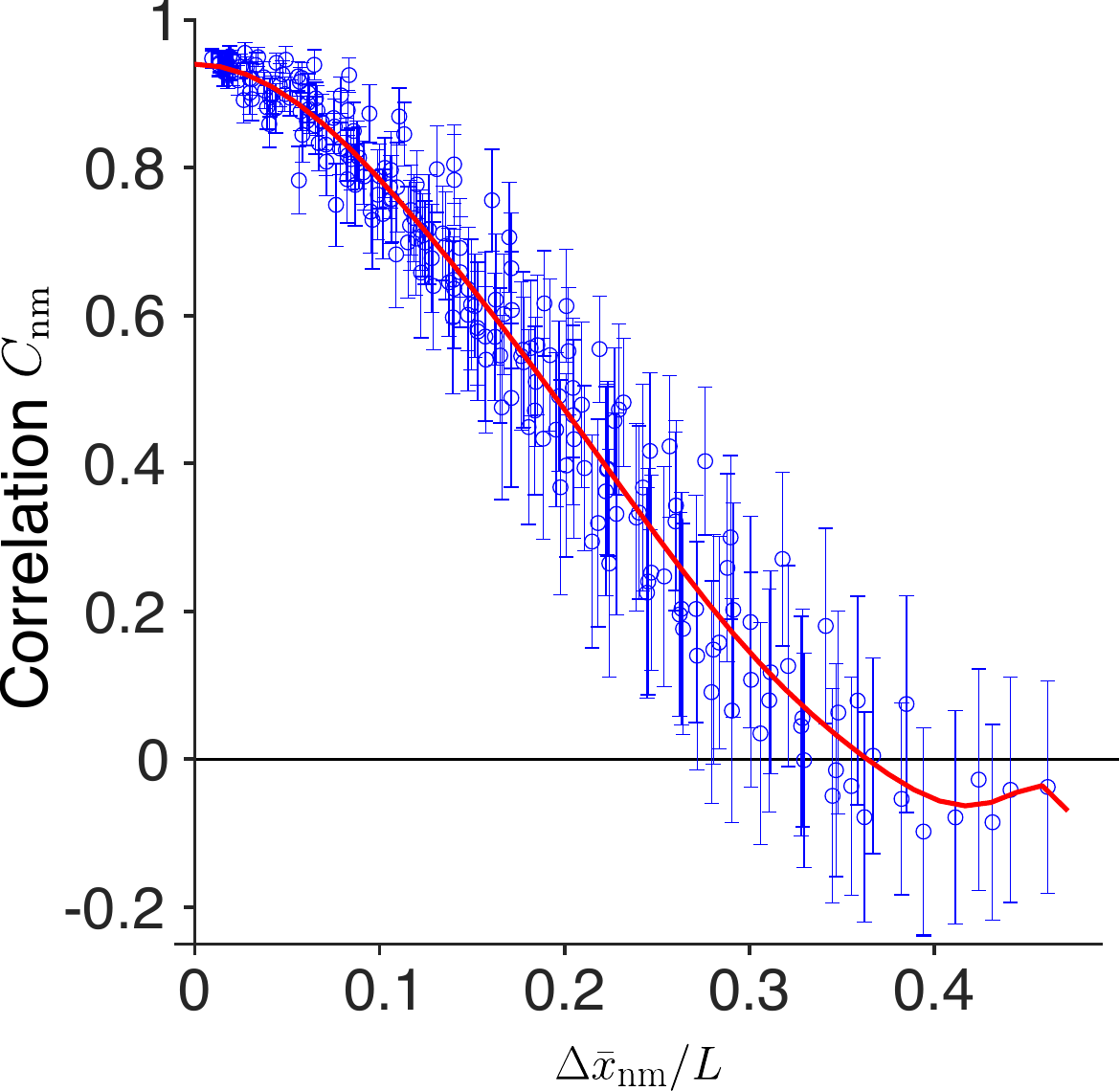}}
\caption{Correlations between noise in peak positions of the {\em eve}, {\em run}, and {\em prd} stripe patterns, as a function of the mean separation between stripes.  Error bars estimated from the standard deviation across random halves of the data.  With three genes, each having seven stripes, we observe  $(21\times 20)/2 = 210$ distinct elements of the correlation matrix $C_{\rm nm}$. Solid red line is a smooth curve to guide the eye. \label{corr1}}
\end{figure}

We are looking at fluctuations in the positions of the stripes, $\delta x_{\rm n}$.  Fig.~\ref{corr1} shows the elements of the correlation matrix
\begin{equation}
C_{\rm nm} \equiv 
{ { \langle\delta x_{\rm n}\delta x_{\rm m}\rangle } \over
{[\langle (\delta x_{\rm n})^2\rangle\langle(\delta x_{\rm m})^2\rangle]^{1/2} } } ,
\end{equation}
as a function of the mean separation $\Delta {\bar x}_{\rm nm}$ between stripes $\rm n$ and $\rm m$.  We see that, within experimental error, the correlations really are a function of distance.  There is no obvious pattern linked to the identity of the enhancers that control these different features, or to the identity of the transcription factors to which the enhancers respond:  nearby stripes are highly correlated, the decay of correlations with distance is the same whether we are looking at correlations between the same or different genes, and different pairs of stripes with same mean separation have the same correlation.   This suggests that, as in the theoretical discussion above, we can think about an abstract positional signal that is transmitted to each cell and controls the placement of the pair-rule stripes.  Correspondingly, there are strong indications that the correlations are inherited from the structure of the noise in gap gene expression (Appendix \ref{gaptopair}).

Qualitatively, the correlations that we see in Fig.~\ref{corr1} decay over distances $\xi \sim 0.15 L$, consistent with the scale needed to close the information gap, and with early measurements \cite{lott+al_07}.  Quantitatively, the decay of correlations is not well described by a single exponential function of distance, so we cannot simply transcribe the predictions of the theory.  Instead, we would like to make a direct estimate of the positional information from the data.  Conceptually this is simple:  we estimate the correlation matrix from the data, then compute the (log) determinant of this matrix following Eq (\ref{eq-DeltaS}).   As with the information gap itself (above), the problem is that random errors in our estimates of individual matrix elements become systematic errors in the entropy.   We follow the same strategy of identifying the dependence of this error on the number of embryos that we include in our analysis and extrapolating to large data sets, as described in Appendix \ref{ent_est}.

\begin{figure}[t]
\centerline{\includegraphics[width = \linewidth]{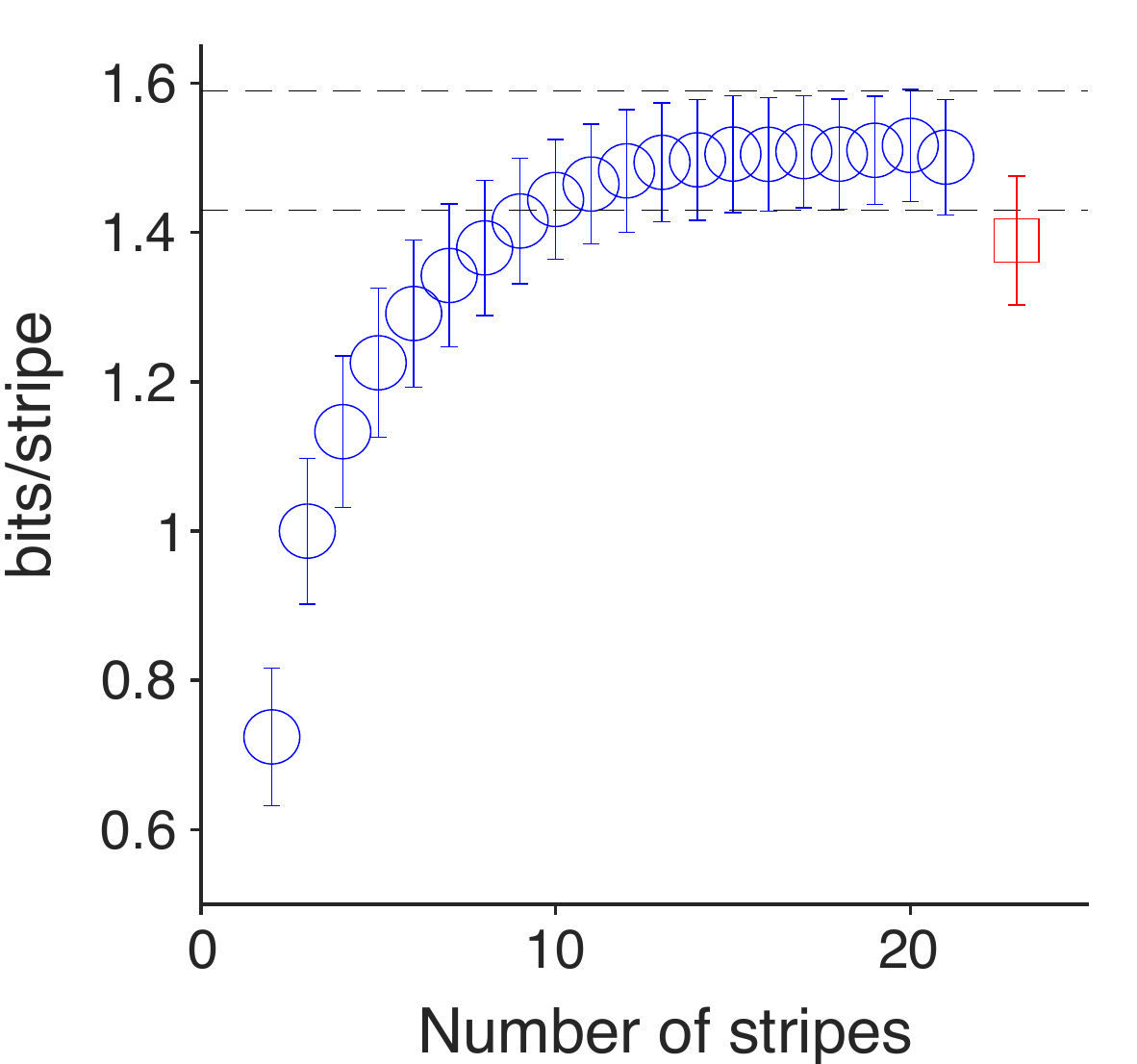}}
\caption{Extra information in correlations per cell, $\Delta S/N$, computed from the observed correlations in pair-rule stripe fluctuations $C_{\rm nm}$, including different numbers of contiguous stripes.   Circles and error bars (blue) are the extrapolated estimates from Appendix \ref{ent_est}.  Beyond $N\sim 14$ stripes there is a plateau $\Delta S/N = 1.51 \pm 0.08\,{\rm bits/cell}$, bracketed by the dashed lines.  Square and error bars (red) are the best estimate of the information gap $I_{\rm gap} = 1.39 \pm 0.08\,{\rm bits/cell}$ from Appendix \ref{one-body}.
\label{I_vs_featnum}}
\end{figure}

By definition, to see the extra information hidden in correlations we have to look at the positions of multiple stripes.  We start with two neighboring stripes, and gradually work out toward all twenty--one stripes.   We see in Fig.~\ref{I_vs_featnum} that beyond $N\sim 10$ stripes, the information per stripe reaches a plateau at $\Delta S/N = 1.51\pm 0.08\,{\rm bits/stripe}$.  This agrees, within experimental error, with our estimate of the information gap $I_{\rm gap} = 1.39 \pm 0.08\,{\rm bits/cell}$.

\section{Discussion}

There is strong evidence that, early in embryonic development,  each cell acquires a distinct identity \cite{gergen+al_86};  it is less clear how this information is encoded. In the fruit fly embryo, positional information along the anterior–posterior axis is orchestrated through a sequential cascade involving three primary maternal inputs, a select number of gap genes, and the pair rule genes. The conventional perspective suggests that the information flow through this cascade entails a gradual refinement, with noisy inputs ultimately generating  a precise and reproducible pattern \cite{arias+hayward_06,lacalli_22}, in the spirit of the Waddington landscape \cite{waddington_57}.

In contrast to the picture of noisy inputs and precise outputs, at least one maternal input itself exhibits a high level of precision, consistently reproducible across embryos \cite{gregor+al_07b,petkova+al_14}. Moreover, the expression levels of gap genes within a single cell prove sufficient to determine positions with an error smaller the distance between neighboring cells \cite{dubuis+al_13,petkova+al_19}. Notably, this precision agrees with that observed in downstream events such as the pair-rule stripes. In parallel, crucial developmental events exhibit highly reproducible temporal trajectories \cite{ferree+al_16}.  These quantitative observations  challenge the conventional view of refinement and error correction, supporting instead a precisionist perspective in which locally available information is processed and preserved with near optimal efficiency.  Given that all relevant molecules are present at low copy numbers, this places significant constraints on the architecture of the underlying networks \cite{ferree+al_16,tkacik+al_09,walczak+al_10,sokolowski+al_23}.

Despite their precision, local signals in the fly embryo do not quite provide enough information to uniquely specify all $N= 74\pm 5$ cellular identities along the AP axis,  $I_{\rm unique} = \log_2 N$:   errors in the  position that a cell can infer from molecular concentrations come from a distribution, and distributions have tails \cite{dubuis+al_13}.  The result is that there is a substantial ($\sim 22\%$) gap between the information provided by the gap genes, or  the pair-rule stripes, and 
$I_{\rm unique}$.

Previous measurements have characterized the noise in local estimates of position for each cell individually.  But there are many hints from previous work that this noise is correlated \cite{lott+al_07,dubuis+al_13,krotov+al_14}.  Extra information can be hiding in these correlations, and we have seen in \S\ref{sec-theory} that if correlations extend over distances $\xi\sim0.15L$ then this would be enough to close the information gap.   This prompts a more detailed examination of the noise correlations, which really do seem to be a function of distance independent of gene identity (Fig.~\ref{corr1}).

The perhaps surprising conclusion of \S\ref{sec-exp} is that the extra information contained in the correlations, $\Delta S/N$, matches the information gap $I_{\rm gap}$ to within a few percent of $I_{\rm unique}$, with the remaining difference essentially equal to our error bars:
\begin{equation}
I_{\rm gap} - \Delta S/N = (-0.019 \pm 0.018)I_{\rm unique}.
\end{equation}
This agreement supports, strongly, the precisionist view of information flow in this system.

Historically, the lack of precise data on gene expression levels, with uncertainties extending to factors of two, led to skepticism regarding the relevance of more refined measurements to general mechanisms of genetic control.  These expectations stood in contrast, for example, to our understanding of signaling in rod photoreceptors, where the quantitative reproducibility of responses to single molecular events provides important constraints on the underlying biochemical mechanisms \cite{doan+al_06}.  

The fly embryo has provided a laboratory within which to explore the precision vs.~noisiness in the function of an intact living system. We have seen reproducible protein and mRNA concentrations across embryos with an accuracy of $10\%$ \cite{gregor+al_07b,dubuis+al_13b,petkova+al_14}, and these concentrations encode position with an accuracy of  $\sim 1\%$ of the embryo's length \cite{dubuis+al_13,petkova+al_19,tkacik+al_15}. The current study adds a layer to this understanding, demonstrating that the available positional information, including the subtle effects of correlated noise, matches the threshold for specifying unique cellular identities, and this match itself has an accuracy of just a few percent.  Beyond the fly embryo, these results suggest a more general conclusion: quantitative measurements in living systems merit serious consideration, even at high precision, as in other areas of physics.

\begin{acknowledgments}
We thank Eric Wieschaus for many inspiring discussions.  This work was supported in part by the US National Science Foundation, through the Center for the Physics of Biological Function (PHY--1734030); by National Institutes of Health Grant R01GM097275;  by the Howard Hughes Medical Institute; and by the Simons and John Simon Guggenheim Memorial Foundations. 
\end{acknowledgments}

\appendix

\renewcommand{\thefigure}{A\arabic{figure}}
\setcounter{figure}{0}

\section{Statistics of individual stripes}
\label{one-body}

The raw data for our analyses are the profiles of fluorescence intensity vs position along the length of the embryo, as in Fig.~\ref{stripes}.   These embryos have been fixed and stained with antibodies against the proteins encoded by the pair rule genes {\em eve}, {\em prd}, and {\em rnt}, and fluorescently tagged antibodies against those antibodies \cite{petkova+al_19}.  Independent experiments demonstrate that these classical staining methods, used carefully, yield fluorescence intensities that are linear in protein concentrations \cite{dubuis+al_13b}. The data set used here, which contains a large number of wild type embryos, comes from Ref \cite{petkova+al_19}. 

\begin{figure}[b]
\includegraphics[width = \linewidth]{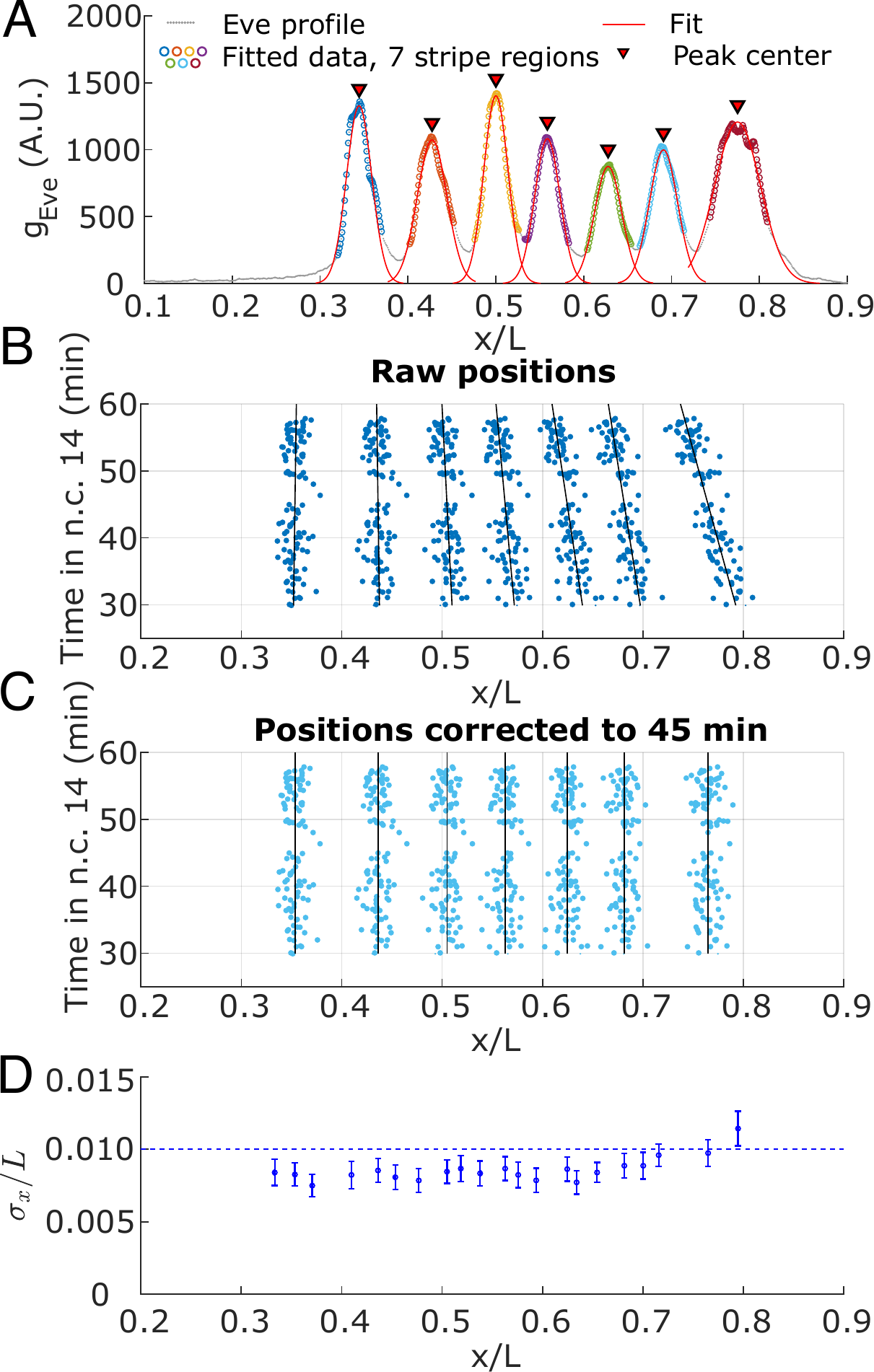}
\caption{Pair rule stripe positions. (A) Concentration of Eve protein in a single embryo. Colored circles indicate regions which were fitted with a Gaussian function to calculate the stripe position. Each stripe is fitted individually, with fits shown in red. Red triangles indicate centers of each fitted peak. (B) Stripe positions as a function of time in the nuclear cycle 14. Linear fits from Eq (\ref{xnvst}) are shown as black lines. (C) Peak positions $x_{\rm n}(t_0)$ corrected to $t_0=45$ min. (D) Positional error of the pair rule stripes.  Magnitude of the error $\sigma_x ({\rm n})$ is plotted against the mean position $\bar x_{\rm n}$ for each of the {\em eve}, {\em prd}, and {\em rnt} stripes.  Errors in $\bar x_{\rm n}$ are standard errors of the mean; errors in $\sigma_x$ are standard deviations across random halves of the data.  Dashed line marks the rough estimate $\sigma_x/L \sim 0.01$. \label{fitting+sigmax}}
\end{figure}

We briefly summarize the imaging protocol and describe the procedure for localizing the stripe positions. Images are taken in the midsaggital plane showing a row of nuclei along the dorsal and ventral side of the embryo. For consistency and to avoid geometric distortion, we focus on the dorsal profiles, as was done previously. In order to include the entire embryos in a single image, large field-of-view images, with pixel size $445\, {\rm nm}$  are acquired with a $20\times$ $0.7$NA objective on a Leica SP5 confocal microscope. Fluorescence intensity is averaged inside a sliding window of the size of a nucleus and the position of the window center is recorded. In a given embryo, positions of the 7 stripes are first roughly identified by finding local maxima in the profile of an individual embryo.   To make this quantitative, we tried several methods.  First, we used an iterative procedure in which the mean peak shape is used as a template \cite{antonetti+al_18}.  Second, we fitted a model of seven Gaussians with variable amplitudes and widths to the entire profile.  Finally, we fit individual Gaussians to each stripe, using a window centered on the local maximum with width of 5\% embryo length.  These methods give consistent results, and importantly global fits do not generate larger correlations than local fits.  In the end we use the local Gaussian fits, as in Fig.~\ref{fitting+sigmax}A.

The age of embryos is estimated to 1 minute precision in nuclear cycle 14 by measuring the length of the cellularization membrane \cite{dubuis+al_13}.  At 30 min into this cycle, the stripes of \emph{prd} first start to become visible and the other two genes have a well defined stripes by that time, so we confine our attention to $t> 30\,{\rm min}$.

Stripe patterns are dynamic, with positions that depend on time.  If we don't take account of this systematic variation, then across an ensemble of embryos with different ages we would see artificial correlations among fluctuations in stripe position.    Stripe movement is small, however, and we can use a linear fit (separately for each of the 21 stripes) across the population of embryos,
\begin{equation}
x_{\rm n}(t) = x_{\rm n}(t_0) + s_{\rm n} (t-t_0).
\label{xnvst}
\end{equation}
Results are shown in Fig.~\ref{fitting+sigmax}B and C. For each embryo we find an equivalent position of all the stripes at a reference time $t_0=45$ min \cite{antonetti+al_18}.

With $x_{\rm n}$ the positions of each pair rule stripe, we have the mean and variance
\begin{eqnarray}
\bar x_{\rm n} &=& \langle x_{\rm n}\rangle\\
\sigma_x^2 ({\rm n}) &=& \langle (x_{\rm n} - \bar x_{\rm n})^2 \rangle ,
\end{eqnarray}
where $\langle \cdots \rangle$ denotes an average over our complete experimental ensemble of $N_{\rm em} = 109$ embryos. Results are shown in Fig.~\ref{fitting+sigmax} D, where we confirm that positional errors are almost all smaller than $1\%$ of the embryo length.

\begin{figure}
\includegraphics[width = \linewidth]{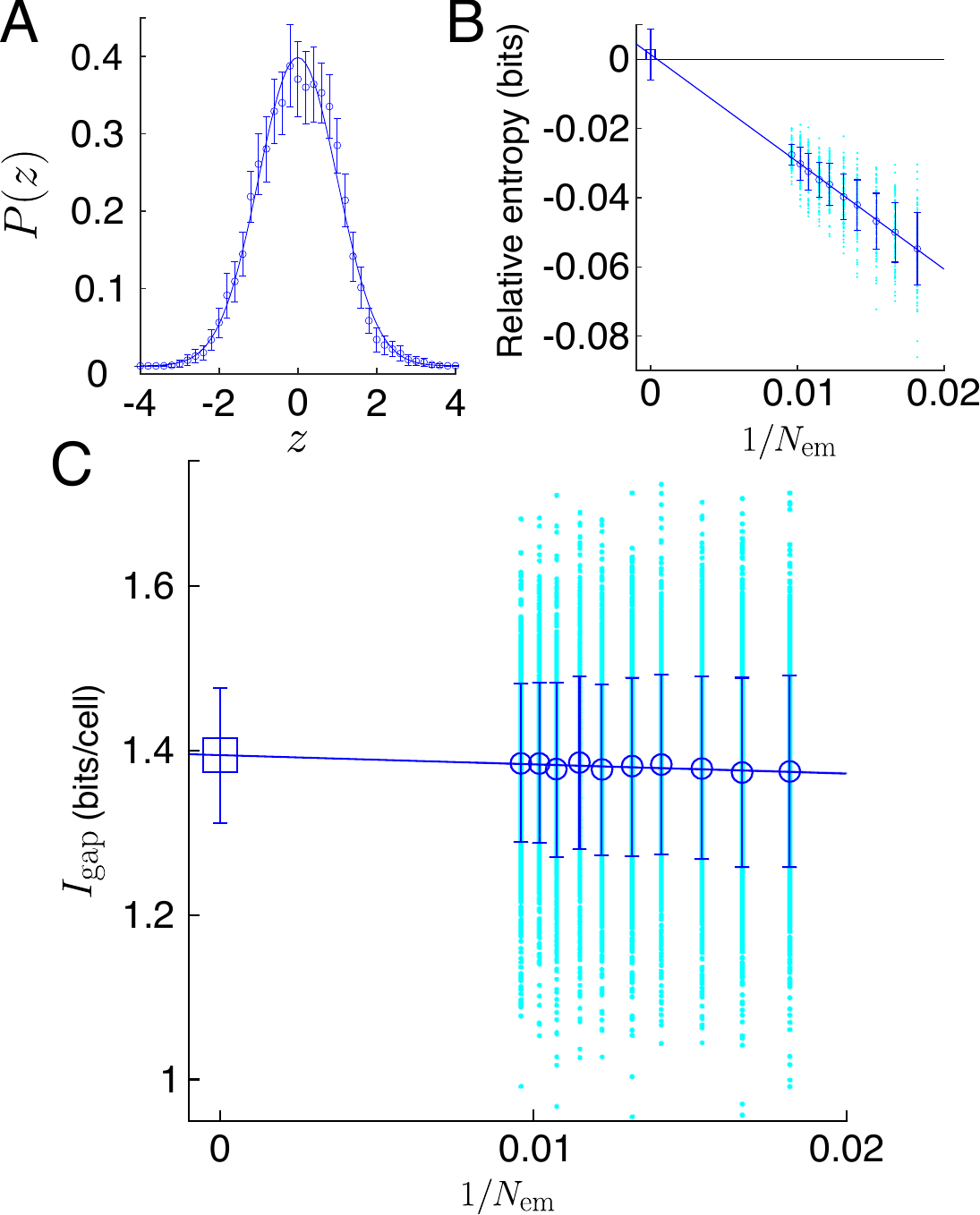}
\caption{(A) Positional errors are well approximated as Gaussian. An estimate of the distribution of normalized errors, Eq (\ref{zscore}).  Open circles are means pooled across all stripes and embryos; error bars are standard deviations across random halves of the embryos; and the line is the Gaussian with zero mean and unit variance. (B) The entropy difference between this estimated distribution and the Gaussian,  as a function of the (inverse) number of embryos we include in our analysis.  Points (cyan) are examples from random choices out of the full ensemble of embryos; open circles with error bars are the mean and standard deviations of these points; and the line is a linear extrapolation \cite{miller_55,panzeri+treves_95,strong+al_98,paninski_03,bialek_12}.  (C) Estimates of the information gap, Eq (\ref{gapA}). Points (cyan) are examples from random choices out of the full ensemble of embryos; open circles (blue) with error bars are the mean and standard deviations of these points; and the line is a linear extrapolation to $I_{\rm gap} = 1.39\pm 0.08 \,{\rm bits}$. \label{gaussianity+Igap} }
\end{figure}

Beyond measuring the variance, we can estimate the distribution of positional errors.  Since the different stripes have slightly different $\sigma_x$, we normalize the positional errors for each stripe individually,
\begin{equation}
z_{\rm n} = (x_{\rm n} - \bar x_{\rm n})/\sigma_x({\rm n}) .
\label{zscore}
\end{equation}
With this normalization we can pool across all 21 stripes, and we estimate the distribution of $z$ as usual by making bins and counting the number of examples in each bin, with results shown at left in Fig.~\ref{gaussianity+Igap}.   Qualitatively the distribution is close to being Gaussian, but what matters for our analysis  is the entropy of this distribution.

When we estimate a probability distribution and use this estimate to compute the entropy, the random errors in the distribution that arise from the finiteness of our sample become systematic errors in the entropy. The general version of this problem goes back to the very first efforts to use information theoretic concepts to analyze biological experiments \cite{miller_55}; for a review see Appendix A.8 of Ref \cite{bialek_12}.   Briefly, naive entropy estimates depend systematically on the size of the sample, and if we can detect this systematic dependence we can extrapolate to infinite data.  At right in Fig.~\ref{gaussianity+Igap} we show the difference between the entropy of the estimated distribution $P(z)$ and the entropy of a Gaussian.  We see that when we base our estimates on $N_{\rm em}$ embryos there is a term $\sim 1/N_{\rm em}$.  Extrapolating $N_{\rm em} \rightarrow\infty$ we see that the entropy difference goes to zero within the small ($< 0.01\,{\rm bit}$) error bars.  We conclude that, for the purposes of our discussion, it is safe to approximate the positional errors as being Gaussian.

Finally we can use the same extrapolation methods to provide a better estimate of the ``information gap'' defined in the main text.  Equation (\ref{Ipos2}) defines the positional information contained in the local signals, $I_{\rm position}$, and the information gap is the difference between this and $I_{\rm unique} = \log_2 N$.  
Fig.~\ref{gaussianity+Igap} shows the values of 
\begin{equation}
I_{\rm gap} = I_{\rm unique} - I_{\rm position} = {\bigg\langle}\log_2\left[\sqrt{2\pi e} {{N\sigma_x({\rm n})}\over L}\right] {\bigg\rangle}_{\rm n}
\label{gapA}
\end{equation}
estimated from fractions of our data set and then extrapolated.  The result is $I_{\rm gap} = 1.39\pm 0.08 \,{\rm bits}$ (Fig.~\ref{gaussianity+Igap}).

\section{Entropy estimates}
\label{ent_est}

\begin{figure}[b]
\includegraphics[width = \linewidth]{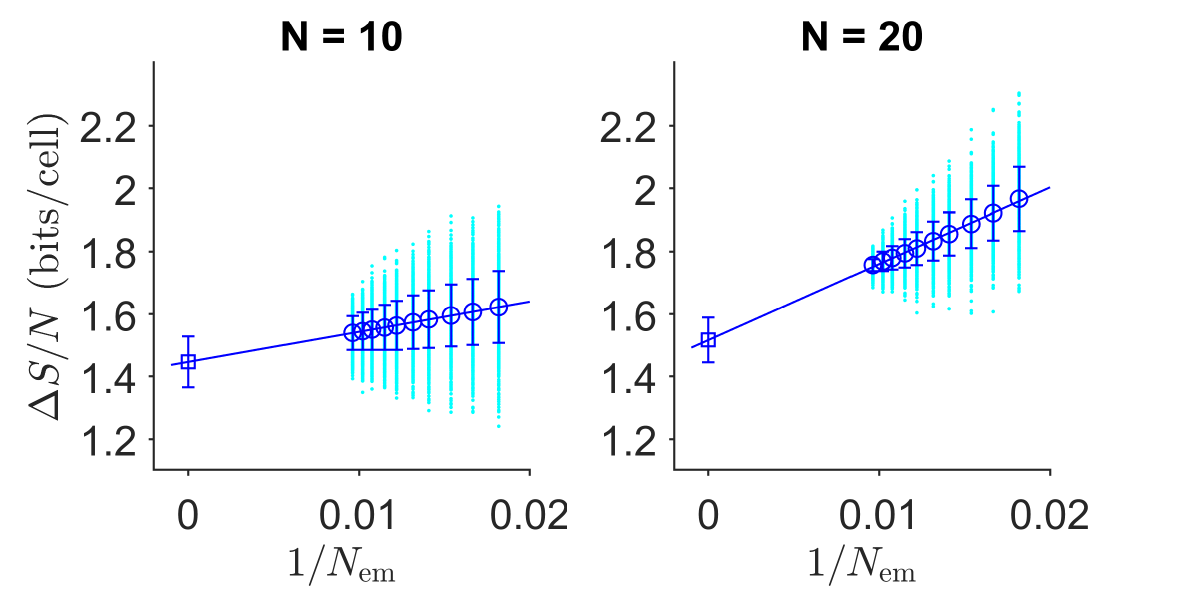}
\caption{Entropy reduction by correlations among the pair rule stripe fluctuations, estimated from different numbers of embryos $N_{\rm em}$; $N=10$ stripes at left and $N=20$ stripes at right.   Points (cyan) are examples from random choices out of the full ensemble of embryos; open circles (blue) with error bars are the mean and standard deviations of these points; and the line is a linear extrapolation to the square. \label{ent_vs_N}}
\end{figure}

Fig.~\ref{ent_vs_N} shows estimates of the extra information $\Delta S/N$ [Eq (\ref{eq-DeltaS})] based on measurements in different numbers of embryos, for $N=10$ and $N=20$ contiguous pair rule stripes.  We see the expected dependence on $1/N_{\rm em}$, and the steepness of this dependence is twice as large at $N=20$ than at $N=10$.  This gives us confidence in the extrapolation $N_{\rm em}\rightarrow\infty$  \cite{miller_55,panzeri+treves_95,strong+al_98,paninski_03,bialek_12}.

\section{Origin of the correlations}
\label{gaptopair}

The precision of pair rule stripe placement matches, quantitatively, the noise in optimal estimates of position based on the local expression levels of the gap genes \cite{dubuis+al_13,petkova+al_19}.  To be consistent with this result, the correlations should also be visible in the gap genes.  As noted above, Lott and colleagues saw correlations in expression boundaries for selected gap genes \cite{lott+al_07}, and later measurements showed that combinations of gap gene expression levels have correlations extending over a significant fraction of the embryo \cite{krotov+al_14}.  Here we revisit these measurements and connect fluctuations in gap gene expression to positional noise.  Notice that for the pair rule genes we can work directly with the positions of the stripes, but for the gap genes we have to think more carefully about how positions are encoded in expression levels.

We start with a brief review of ideas about decoding positional information \cite{petkova+al_19}.
Measurements of gap gene expression in multiple embryos provide samples from the conditional distribution $P(\{g_{\rm i}\}|x)$, at all values of the position $x$ along the anterior--posterior axis; we focus on the $d=4$  gap genes expressed in the middle $\sim 80\%$ of the embryo, {\em hunchback}, {\em giant}, {\em kr\"uppel}, and {\em knirps}. To a good approximation this distribution is Gaussian,
\begin{eqnarray}
P(\{g_{\rm i}\}|x) &=&{1\over {Z(x)}} \exp\left[ - {1\over 2}\chi^2\left( \{g_{\rm i}\} ; x\right)\right]\label{Pgx1}\\
Z(x) &=&   \left[ (2\pi)^d \det\hat C (x)\right]^{1/2}\\
\chi^2\left( \{g_{\rm i}\} ; x\right) &=& \sum_{{\rm i,j}=1}^d \left[ g_{\rm i} - \bar g_{\rm i}(x)\right] \left[\hat C^{-1}(x)\right]_{\rm ij} \left[ g_{\rm j} - \bar g_{\rm j}(x)\right]  ,\nonumber\\
&& 
\end{eqnarray}
where $\bar g_{\rm i}(x)$ is the mean expression level of gene $\rm i$ at position $x$ and 
\begin{equation}
\left[\hat C (x)\right]_{\rm ij} = \langle \delta g_{\rm i} \delta g_{\rm j}\rangle_x
\end{equation}
is the covariance matrix of fluctuations around these means.   To decode the position of a cell from the local expression levels we need to construct 
\begin{equation}
P\left(x | \{g_{\rm i}\}\right) = {{P(\{g_{\rm i}\}|x) P(x)}\over{P(\{g_{\rm i}\})}} .
\end{equation}
But because nuclei are arrayed uniformly along the length of the embryo, $P(x)$ is uniform and hence the dependence on $x$ is captured in Eq (\ref{Pgx1}).  

 \begin{figure}[b]
\centerline{\includegraphics[width = \linewidth]{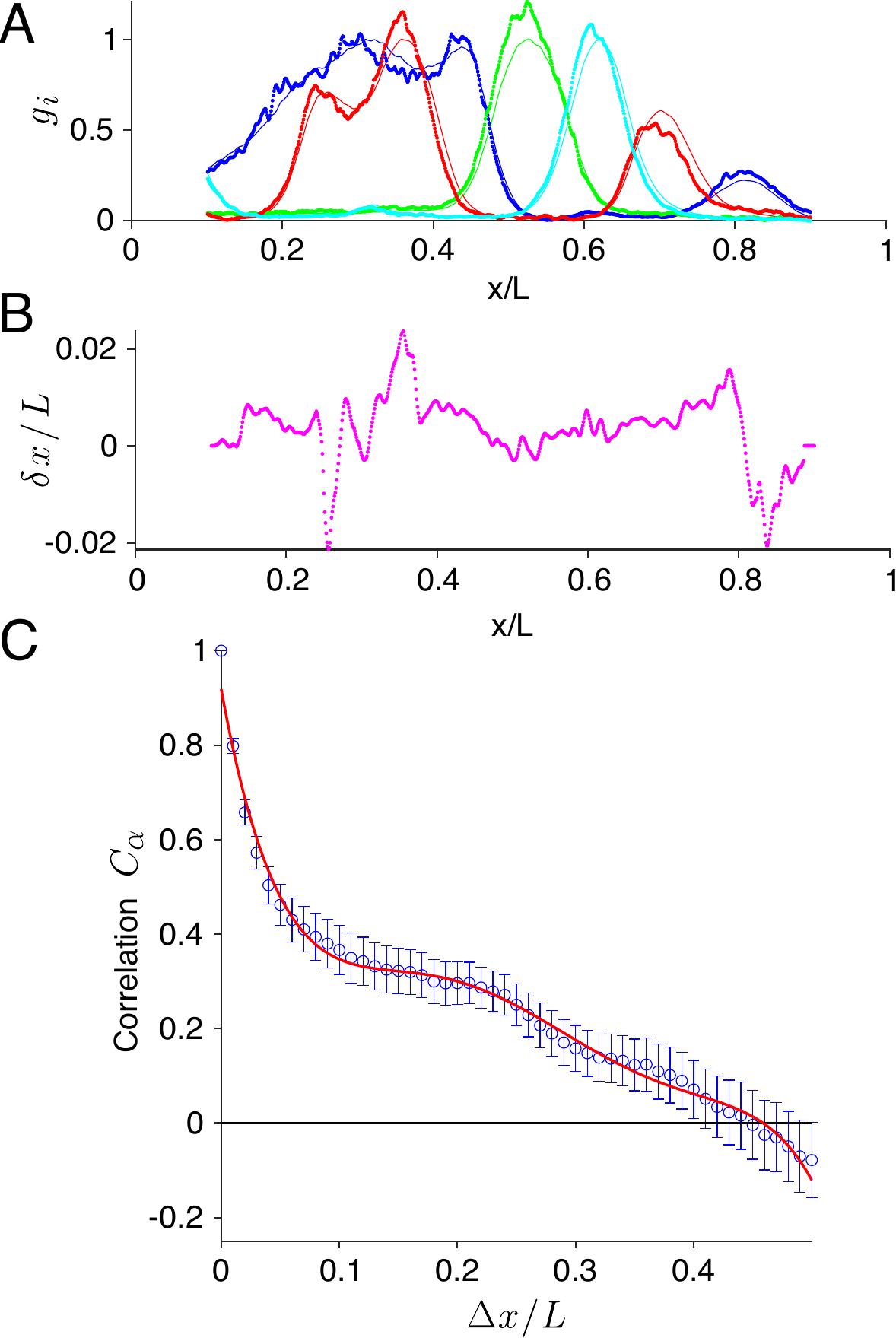}}
\caption{Decoding gap gene expression levels in a single embryo and correlations in the resulting pattern of positional errors.  (A) Expression of Hb (blue), Kr (green), Gt (cyan), and Kni (red).  Thin solid lines are means across $N_{\rm em} = 38$ embryos in a small window $40 \leq t \leq 44\,{\rm min}$ in nuclear cycle 14; dense points are data from a single embryo \cite{petkova+al_19}.  (B) Positional errors computed from Eq (\ref{deltax_final}).  (C) Correlations in the positional noise inferred from gap gene expression.  For each embryo $\alpha$ we compute the correlation function in Eq (\ref{corfn}) and then normalize to give $\tilde C (\Delta x) =  C (\Delta x)/C (0)$.  Blue circles with error bars are mean and standard error across $N_{\rm em} = 38$ embryos; solid red line is a smooth curve to guide the eye.  
 \label{gapdxfig+gap_corr}}
\end{figure}

A cell at the actual position $x_{\rm true}$ has expression levels
\begin{equation}
g_{\rm i} = \bar g_{\rm i}(x_{\rm true}) + \delta g_{\rm i} ,
\end{equation}
and if the positional noise is small we can write
\begin{equation}
\bar g_{\rm i} (x) = \bar g_{\rm i}(x_{\rm true}) + (x - x_{\rm true}) {{d\bar g_{\rm i}(x)}\over{dx}} {\bigg |}_{x=x_{\rm true}} + \cdots.
\end{equation}
If the noise is small, then the the best estimate of position based on the gap gene expression levels is the value of $x$ which minimizes $\chi^2$, and this can be written as
\begin{eqnarray}
\hat x  &=& x_{\rm true} + \delta x\\
\delta x (x_{\rm true}) &=& \left[\sigma_x^2 (x) \sum_{{\rm i,j}=1}^d \delta g_{\rm i} \left[\hat C^{-1}(x)\right]_{\rm ij} {{d\bar g_{\rm j}(x)}\over{dx}}\right]_{x = x_{\rm true}} ,
\nonumber\label{deltax_final}\\
&&
\end{eqnarray}
where the variance of positional noise is defined by
\begin{equation}
{1\over{\sigma_x^2(x)}} = \sum_{{\rm i,j}=1}^d {{d\bar g_{\rm i}(x)}\over{dx}} \left[\hat C^{-1}(x)\right]_{\rm ij}  {{d\bar g_{\rm j}(x)}\over{dx}} ;
\end{equation}
for consistency we have
\begin{equation}
\langle [\delta x (x)]^2\rangle = \sigma_x^2(x) .
\end{equation}
Note the connection to Eqs (\ref{est1}) and (\ref{est2}) in \S\ref{sec:problem}.

Previous work has emphasized the scale of positional errors $\sigma_x$ \cite{dubuis+al_13,tkacik+al_15,petkova+al_19}.  But the optimal decoding of gap gene expression levels \cite{petkova+al_19}  maps the deviation of expression levels from the mean into a decoding error for each embryo individually, as in Eq (\ref{deltax_final}).  An example  is  in Fig.~\ref{gapdxfig+gap_corr}, where the small fluctuations of expression levels around the mean (A) translate into proportionally small errors $\delta x$ (B).

For each embryo $\alpha$ we can take the positional errors $\delta x_\alpha (x)$ and compute the correlation function
\begin{equation}
C_\alpha (\Delta x) = {1\over{L-\Delta x}}\int dx\, \delta x_\alpha (x) \delta x_\alpha (x+\Delta x) .
\label{corfn}
\end{equation}
Fig.~\ref{gapdxfig+gap_corr}C shows the mean and standard error of the normalized correlation function across all $N_{\rm em } = 38$ embryos in our experimental ensemble.  Qualitatively, correlations in the positional noise encoded by the gap genes extend over distances similar to the correlation in positional noise of the pair rule stripes (Fig.~\ref{corr1}).  Quantitatively, the gap gene correlations include an additional component with a short correlation length.  One possibility is that this component is averaged away by interactions among neighboring cells during expression of the pair rule stripes.  Another possibility is that a modest fraction of the noise in gap gene expression reflects local noise in the measurements, as discussed previously \cite{dubuis+al_13b}; this measurement noise has only a small impact on our estimates of the effective noise $\sigma_x$ but a larger impact on the shape of the correlation function.  It seems likely that both effects contribute.  Nonetheless, it is clear that relatively long ranged correlations, which are crucial to closing the information gap, are present already in the gap gene expression levels, as suggested in earlier work \cite{lott+al_07,dubuis+al_13,krotov+al_14}.  New experiments will be needed to give a reliable estimate of the information that is encoded in these correlations.


\begin{thebibliography}{99}
%
\bibitem{turing_52}
AM Turing,   The chemical basis of morphogenesis. {\em Philos Trans R Soc Lond  B, Biol Sci} {\bf  237,} 37--71 (1952).
%
\bibitem{wolpert_69}
L Wolpert,  Positional information and the spatial pattern of cellular
differentiation. {\em J Theor Biol}  {\bf 25,} 1--47 (1969). 
%
\bibitem{tkacik+gregor_21}
G Tka\v{c}ik and T Gregor, The many bits of positional information.  {\em Development} {\bf 148,} dev176065 (2021). 
%
\bibitem{nusslein+wieschaus_80}
C Nusslein--Vollhard and E Wieschaus, Mutations affecting segment number and polarity in {\em Drosophila}. {\em Nature} {\bf 287,} 795--801 (1980).
%
\bibitem{lawrence_92}
PA Lawrence. {\em The Making of a Fly: The Genetics of Animal Design} (Blackwell Scientific, Oxford, 1992).
%
\bibitem{ew+cnv_16}
E Wieschaus and C N\"usslein--Volhard, The Heidelberg screen for pattern mutants of {\em Drosophila}: A personal account. {\em Annu Rev Cell Dev Biol} {\bf 32,} 1--46 (2016).
%
\bibitem{riverapomar_96}
R Rivera--Pomar and H Jackle,  From gradients to stripes in {\em Drosophila} embryogenesis: Filling in the gaps. {\em Trends Genet} {\bf 12,} 478--483 (1996).
%
\bibitem{jaeger_11}
J Jaeger, The gap gene network. {\em Cell Mol Life Sci} {\bf 68,} 243--274 (2011).
%
\bibitem{gergen+al_86}
JP Gergen, D Coulter, and EF Wieschaus, Segmental pattern
and blastoderm cell identities. In {\em Gametogenesis and The Early
Embryo}, JG Gall, ed, pp 195--220 (Liss, New York, 1986).
%
\bibitem{bugs}
See https://bugguide.net/node/view/1308194/bgimage. 
%
\bibitem{dubuis+al_13}
JO Dubuis, G Tka\v{c}ik, EF Wieschaus,   T Gregor, and W Bialek, Positional information, in  bits. {\em Proc Natl Acad Sci (USA)} {\bf 110,}  16301--16308 (2013).
%
\bibitem{liu+al_13}
F Liu, AH Morrsison, and T Gregor,  Dynamic interpretation of maternal inputs by the {\em Drosophila} segmentation gene network.  {\em Proc Natl Acad Sci (USA)} {\bf 110,}  6724--6729 (2013).
%
\bibitem{petkova+al_19}
MD Petkova, G Tka\v{c}ik, W Bialek, EF Wieschaus, and T Gregor,  Optimal decoding of cellular identities in a genetic network.   {\em Cell} {\bf 176,} 844--855 (2019).
%
\bibitem{yu+al_06}
J Yu, J Xiao, X Ren, K Lao, and XS Xie, Probing gene expression in live cells, one protein molecule at a time. {\em Science} {\bf 311,} 1600--1603 (2006).
%
\bibitem{taniguchi+al_10}
Y Taniguchi, PJ Choi, G--W Li, H Chen, M Babu, J Hearn, A Emili, and XS Xie, Quantifying {\em E.~coli} proteome and transcriptome with single--molecule sensitivity in single cells. {\em Science} {\bf 329,} 533--538 (2010).
%
\bibitem{dubuis+al_13b}
JO Dubuis, R Samanta, and T Gregor, Accurate measurements of dynamics and reproducibility in small genetic networks.
{\em Mol Sys Biol} {\bf 9,} 639 (2013).
%
\bibitem{lott+al_07}
SE Lott, M Kreitman, A Palsson, E Alekseeva, and MZ Ludwig, Canalization of segmentation and its evolution in {\em Drosophila}. {\em Proc Natl Acad Sci (USA)} {\bf 104,} 10926--10931 (2007).
%
\bibitem{krotov+al_14}
 D Krotov, JO Dubuis, T Gregor, and W Bialek, Morphogenesis at criticality.  {\em Proc Natl Acad Sci (USA)} {\bf 111,} 3683--3688 (2014).
%
\bibitem{nusslein-volhard_89}
C Nüsslein-Volhard  and S Roth, Axis determination in insect embryos. {\em Ciba Found Symp} {\bf 144,} 37--55 (1989).
%
\bibitem{nusslein-volhard_91}
C Nüsslein-Volhard, Determination of the embryonic axes of Drosophila. {\em Dev Suppl} {\bf 1,} 1-10 (1991).
%
\bibitem{tkacik+al_15}
G Tka\v{c}ik, JO Dubuis,  MD Petkova, and T Gregor,
Positional information, positional error, and read--out precision in morphogenesis: a mathematical framework.  {\em Genetics} {\bf 199,} 39--59 (2015). 
%
\bibitem{shannon_48}
CE Shannon, A mathematical theory of communication.
{\em Bell Sys Tech J}  {\bf 27,} 379--423 and 623--656 (1948).
%
\bibitem{bialek_12}
W Bialek, {\em Biophysics: Searching for Principles} (Princeton University Press, Princeton NJ, 2012).
 %
\bibitem{gregor+al_07b}
T Gregor, DW Tank, EF Wieschaus, and W Bialek, Probing the limits to positional information.   {\em Cell} {\bf 130,} 153--164 (2007).
%
\bibitem{antonetti+al_18}
V Antonetti,  W Bialek,  T Gregor,  G Muhaxheri, M Petkova, and M Scheeler,  Precise spatial scaling in the early fly embryo.  arXiv:1812.11384 [q--bio.MN] (2018).
%
\bibitem{miller_55}
GA Miller, Note on the bias of information estimates.  In {\em Information Theory in
Psychology: Problems and Methods II-B}, H Quastler, ed., pp. 95–100 (Free Press, Glencoe IL,
1955).
%
 \bibitem{panzeri+treves_95}
 S Panzeri and A Treves, The upward bias in measures of information derived from limited data samples.  {\em Neural Comp} {\bf 7,} 399--407 (1995).
 %
\bibitem{strong+al_98}
SP Strong, R Koberle, RR de Ruyter van Steveninck, and W Bialek, Entropy and information in neural spike trains.   {\em Phys Rev Lett} {\bf 80,} 197--200 (1998).
%
\bibitem{paninski_03}
L Paninski, Estimation of entropy and mutual information. {\em Neural Comp} {\bf 15,} 1191--1253 (2003).
%
\bibitem{arias+hayward_06}
AM Arias and P Hayward, Filtering transcriptional noise during development: concepts and mechanisms. {\em Nat Rev Genet} {\bf 7,} 34--44 (2006).
%
\bibitem{lacalli_22}
TC Lacalli, Patterning, from conifers to consciousness: Turing’s theory and order from fluctuations.  {\em Front  Cell  Dev  Biol} {\bf 10,} 871950 (2022).
%
\bibitem{waddington_57}
CH Waddington, {\em The Strategy of the Genes. A Discussion of Some Aspects of Theoretical Biology.}  (Allen and Unwin, London, 1957).
%
\bibitem{petkova+al_14}
MD Petkova, SC Little, F Liu, and T Gregor, Maternal origins of developmental reproducibility. {\em Curr Biol} {\bf 24,} 1283--1288 (2014).
%
\bibitem{ferree+al_16}
PL Ferree, VE Deneke, and S Di Talia,  Measuring time during early embryonic development. {\em Semin Cell Dev Biol} {\bf 55,} 80--88  (2016).
%
\bibitem{tkacik+al_09}
 G Tka\v{c}ik, AM Walczak, and W Bialek, Optimizing information flow in small genetic networks.  {\em Phys Rev E} {\bf 80,} 031920 (2009).
 %
\bibitem{walczak+al_10}
AM Walczak, G Tka\v{c}ik, and W Bialek, Optimizing information flow in small genetic networks. II: Feed--forward interaction.   {\em Phys Rev E} {\bf 81,} 041905 (2010).
%
\bibitem{sokolowski+al_23}
TR Sokolowski, T Gregor, W Bialek, and G Tka\v{c}ik,  Deriving a genetic regulatory network from an optimization principle.   arXiv:2302.05680 [physics.bio--ph] (2023).
 %
\bibitem{doan+al_06}
T Doan, A Mendez, PB Detwiler, J Chen, and F Rieke, Multiple phosphorylation sites confer reproducibility of the rod's single-photon responses. {\em Science} {\bf 313,} 530--533 (2006).
%
\end{thebibliography}
\end{document}